\documentclass[aps,reprint,superscriptaddress]{revtex4-1}
\usepackage{amsmath,amsfonts,amssymb}
\usepackage{graphicx,float}
\usepackage{epstopdf}

\usepackage[normalem]{ulem}
\usepackage{color}
\newcommand{\note}[1]{\textcolor{red}{#1}}

\newcommand{\kpr}{k^{\prime}}
\newcommand{\xpr}{x^{\prime}}
\newcommand{\avg}[1]{\left\langle #1 \right\rangle}

\begin{document}
%\begin{CJK*}{UTF8}{bkai}
\title{The Transsortative Structure of Networks}
\author{Xin-Zeng Wu}% (吳信增)} % (\begin{CJK}{UTF8}{bkai}吳信增\end{CJK})
\affiliation{Information Sciences Institute, University of Southern California, Marina del Rey, CA 90292}
\affiliation{Department of Physics and Astronomy, University of Southern California, Los Angeles, CA 90089}
\author{Allon G.~Percus}
\affiliation{Institute of Mathematical Sciences, Claremont Graduate University, Claremont, CA 91711}
\affiliation{Information Sciences Institute, University of Southern California, Marina del Rey, CA 90292}
\author{Keith Burghardt}
\email{keithab@isi.edu}
\affiliation{Information Sciences Institute, University of Southern California, Marina del Rey, CA 90292}
\author{Kristina Lerman}
\affiliation{Information Sciences Institute, University of Southern California, Marina del Rey, CA 90292}

\begin{abstract}

%%%% Keith's version of the abstract
Network topologies can be non-trivial, due to the complex underlying behaviors that form them. % Despite this, past research has shown that some processes on networks can be characterized by low-order statistics, such as degree assortativity, that describe %correlations between
% nodes and their neighbors. 
%%% Introduce tension
While past research has shown that some processes on networks may be characterized by low-order statistics describing nodes and their neighbors, such as degree assortativity, these quantities fail to capture important sources of variation in network structure.
We introduce a property called transsortativity that describes correlations among a node's neighbors, generalizing these statistics from immediate one-hop neighbors to two-hop neighbors. We describe how transsortativity can be systematically varied, independently of the network's degree distribution and assortativity. Moreover, we show that it can significantly impact the spread of contagions as well as the perceptions of neighbors, known as the majority illusion. Our work improves our ability to create and analyze more realistic models of complex networks.

%{\bf One Sentence Summary}\\
%Our work demonstrates how transsortativity, which captures correlations between neighbors of a node, can improve our understanding of network dynamics.
\end{abstract}

\maketitle
%\end{CJK*}

\section{Introduction}

Networks serve as a substrate for the spread of contagion %contagious outbreaks
in social groups~\cite{Watts2002}, propagation of information in online platforms~\cite{Lerman2016meme}, and cascading failures in the electrical power grid as well as in the financial sector~\cite{Watts1998,PastorSatorras2015,Newman2002Disease,Bianconi2002,Dorogovtsev2008,Goh2003,Goltsev2008,Lin2018,Turalska2019}. Networks are frequently modeled using random graphs~\cite{Newman2566,Noldus2015,Molloy1995,Hofstad2016} that preserve certain statistical properties of real networks, such as degree distribution or degree assortativity~\cite{Newman2002Assort}, while removing other structure. These random graph models (RGMs) have been critical to understanding phenomena such as percolation, disease propagation, and ferromagnetism~\cite{PastorSatorras2015,Newman2002Disease,Bianconi2002,Dorogovtsev2008,Goh2003,Dodds2009,Payne2009,Wu2018}.
However, by ignoring some of the variations inherent in real-world networks, such RGMs offer an incomplete or inaccurate understanding of network phenomena~\cite{Peel2018,Fowler1720}. %Specifically, networks often have higher-order structure for several reasons, including arising from many-body interactions beyond dyadic interactions of pairs of nodes~\cite{benson2018simplicial}.
%As a result, they
Networks, for example, have far more connected triplets and larger motifs than RGMs would typically predict~\cite{ugander2013subgraph}. Also, the neighbors of a node can be similar to one another even when they are not similar to the node itself~\cite{Altenburger2018}. Such higher-order and longer-range structure has proven instrumental in explaining effects such as the strong friendship paradox~\cite{Wu2017}, where the majority of a node's neighbors have higher degree than the node itself~\cite{Kooti2014}.
% can be quite different from the node’s own attributes, is often observed in real-world networks.  It is a necessary
%Often random graph models do not account for such structure, but what effect this has is not well understood.%thereby limiting their utility as a tool in the study of networks.

We describe a method to measure and model %longer-range
higher-order network structure that offers a more complete description of network phenomena. %Just as assortativity inspired work by Newman on degree assortativity \cite{Newman2002Assort}, we discover that current RGMs are inadequate to accurately describe correlations between individual that are not directly connected~\cite{Altenburger2018}.
We introduce the notion of transsortativity as a measure of degree correlations among a node's neighbors (two-hop) and illustrate how it can significantly alter network structure and network phenomena. Namely, we show that transsortativity amplifies the ``majority illusion'' effect, where an unpopular idea may be perceived as popular by a large fraction of individuals, and impacts the size and critical threshold for cascades in the Watts threshold model~\cite{Watts2002}. %Moreover, it can accurately capture --- and generalize---the two-hop correlations found in a paper by Altenburger et al.~\cite{Altenburger2018}.  % by {\color{red}{reducing the critical threshold at which cascades spread globally}}. %increasing the maximum critical threshold needed to prevent the spread of contagion}}.
We show that the metric helps generalize and provide an explanation for overdispersion (monophily) in social networks~\cite{Zheng2006,Altenburger2018}, where the attributes of a node's neighbors display a larger variance than expected, and also explains the less familiar case of underdispersion. Finally, we describe a rewiring procedure to systematically vary transsortativity while keeping fixed the %more basic
lower-order structure of a network, namely its degree distribution and assortativity. Our examples demonstrate that transsortativity is an important tool in the statistical modeling of networks, and can be an essential extension to configuration models~\cite{Molloy1995,Hofstad2016} and random rewiring~\cite{Noldus2015}, enabling more accurate predictions on realistic networks.

\section{Results}
\subsection{Quantifying Transsortativity in Networks}
Our analysis is motivated by the
% Before we define a metric for neighbor assortativity, we will first discuss the simpler assortativity metric. The
\emph{dK}-series of probability distributions~\cite{Mahadevan2006},
% The \emph{dK}-series
which specifies the joint distribution of the degrees of connected subgraphs of \emph{d} nodes. This provides a useful framework for characterizing network structure. The degree distribution of a network, $p(k)$, represents its \emph{1K} structure.
%, since it considers each node's degree on its own.
The joint degree distribution of pairs of adjacent nodes,
% This distribution is specified by the joint degree matrix
$e(k,\kpr)$, represents its \emph{2K} structure.
% whose elements represent the probability of finding an edge connecting nodes with degrees $k$ and $\kpr$.
% The \emph{2K} distribution gives
The Pearson correlation coefficient of the degrees of a node and of its neighbor is known as the \emph{degree assortativity}~\cite{Newman2002Assort}:
\begin{align}
r_{2K} &= \frac{\mathrm{Cov}(k,k')}{\mathrm{Var}(k)} \nonumber \\
&= \frac{\sum_{k,{\kpr}}k\kpr\left[e(k,{\kpr})-q(k)q({\kpr})\right]}{\sum_k k^2 q(k) - \left[\sum_k k q(k)\right]^2},
\label{assort}
\end{align}
where $q(k)=\sum_{\kpr}e(k,\kpr)=kp(k)/\avg{k}$ is the degree distribution of a node that is adjacent to another, and $\mathrm{Cov}(k,k')$ and $\mathrm{Var}(k)$ are taken with respect to $q(k)$.

Now consider the neighbors of a degree-$k$ node.  Their degree distribution is
% of a random neighbor of a degree-$k$ node
$\nu(\kpr|k)%=P(\kpr|k)
=e(k,\kpr)/q(k)$.
% This distribution has mean $\kpr_{2K}(k)=$\mathrm{E}[\kpr|k]$ and variance %$\sigma^2_{2K}(k)=$\mathrm{Var}(\kpr|k)$.
In many real-world networks, given a pair of such neighbors $i$ and $j$, one finds that their degrees $\kpr_i$ and $\kpr_j$ are %positively
correlated even if $i$ and $j$ are not themselves linked by an edge~\cite{Wu2017}. This two-hop degree correlation reflects the higher-order network structure, specifically the \emph{3K} structure characterizing connected subgraphs with three nodes forming a wedge or a triangle.

Let $w(\kpr_i,\kpr_j|k)$ %=t(\kpr_i,k,\kpr_j)/\nu(k)
denote the joint degree distribution for those two neighbors.  We define the correlation coefficient of $\kpr_i$ and $\kpr_j$ as
%We define the correlation coefficient of neighbors of a degree-$k$ node as
\begin{align}
r_{3K}(k)&=\frac{\mathrm{Cov}(\kpr_i,\kpr_j|k)}{\mathrm{Var}(\kpr|k)} \nonumber \\
&=\frac{\sum_{\kpr_i,\kpr_j}\kpr_i\kpr_j\left[w(\kpr_i,\kpr_j|k)-\nu(\kpr_i|k)\nu(\kpr_j|k)\right]}{\sum_{\kpr} (\kpr)^2 \nu(\kpr|k)-[\sum_{\kpr}\nu(\kpr|k)]^2},
\label{nbcorr}
%\label{nbcorr}
\end{align}
where $\mathrm{Cov}(\kpr_i,\kpr_j|k)$ and $\mathrm{Var}(\kpr|k)$ are taken with respect to $\nu(\kpr|k)$. %, and $w(\kpr_i,\kpr_j|k)=t(\kpr_i,k,\kpr_j)/\nu(k)$
%is the joint degree distribution of two neighbors of a degree-$k$ node.
We refer to $r_{3K}(k)$ as \emph{transsortativity}, because it measures correlations \emph{across} neighbors rather than between a node and its neighbor. Transsortativity generalizes the notion of assortativity from immediate, or one-hop, neighbors to two-hop neighbors.

Values of transsortativity are bounded. To see why, consider the mean degree of a neighbor of a degree-$k$ node, $\bar{k}'=\sum_i \kpr_i/k$.  The variance of this quantity is
\begin{align}\label{3kvar}
\mathrm{Var}(\bar{k}'|k)
&=\frac{1}{k^2}\left[\sum_{i=1}^k\mathrm{Var}(\kpr_i|k)+2\sum_{i=1}^{k-1}\sum_{j=i+1}^{k}\mathrm{Cov}(\kpr_i,\kpr_j|k)\right]\nonumber\\
% &=\frac{1}{k^2}\left[k\mathrm{Var}(\kpr|k)+k(k-1)\;\mathrm{Cov}(\kpr_i,\kpr_j|k)\right]\nonumber\\
% &=\frac{1}{k}\sigma^2_{2K}(k)[1+(k-1)r_{3K}(k)]
&= \frac{\mathrm{Var}(\kpr|k)}{k} [1+(k-1)r_{3K}(k)].
\end{align}
Nonnegativity of the variance gives the lower bound:
\begin{equation}
r_{3K}(k)\ge -\frac{1}{k-1}.
\label{nbcorr_lb}
\end{equation}

\begin{figure}
  \centering
\includegraphics[width=\columnwidth]{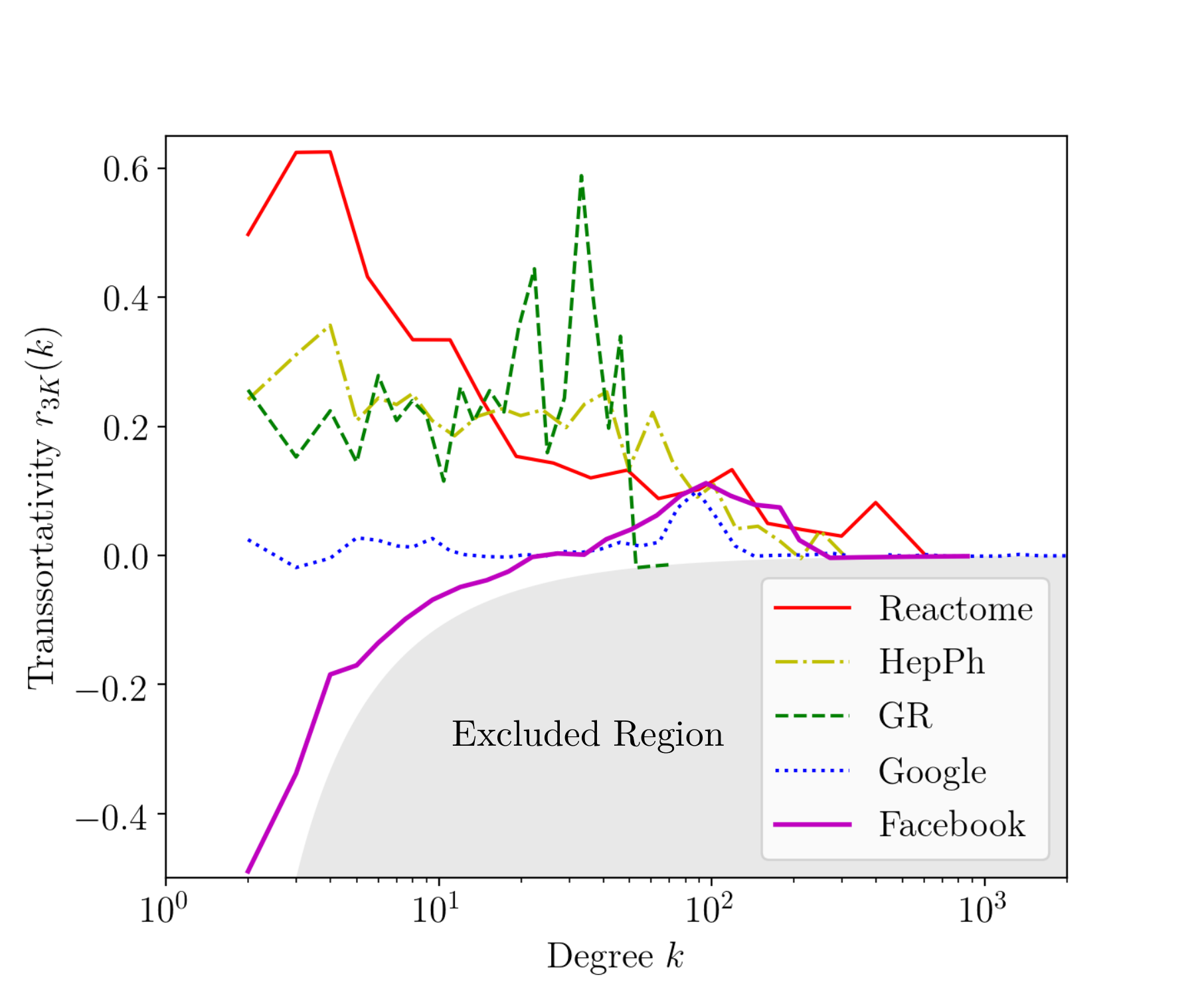}%nbcorr_newtext-crop_Mathematica.pdf}%{nbcorr.eps}
  \caption{(Color online) Transsortativity
% $\bar{r}_{3K}$ from degree specific
$r_{3K}(k)$ for networks from a variety of domains. The networks~\cite{snapnets} include biological (Reactome), co-authorship (HepPh, GR), technological (Google), and social (Facebook). See Supplemental Material for further examples. Gray region shows transsortativity values excluded by the theoretical lower bound (Eq.~\eqref{nbcorr_lb}). Data is aggregated using log-binning on degree $k$.
}\label{fig:nbcorr}
\end{figure}

%\subsection{Real-World Networks}
Examples of transsortativity in real-world networks~\cite{snapnets} are given in Fig.~\ref{fig:nbcorr}, showing that observed values of $r_{3K}(k)$ are large in cases ranging from a biological network of protein-protein interactions (Reactome), to co-authorship networks between physicists (ArXiv HepPh and GR), to hyperlink networks between webpages (Google), to friendship social networks (Facebook).  Note that in most of these networks, transsortativity values are  %not only bounded by Eq.~(\ref{nbcorr_lb}) but are
positive, implying assortative mixing between two-hop neighbors, regardless of the degree assortativity of immediate neighbors. %One notable exception is the
Surprisingly, the Facebook social graph exhibits substantially negative transsortativity for low-degree nodes. This implies that low-degree nodes are connected to both low-degree and high-degree neighbors.

By averaging over all degrees in the network, we can calculate the mean transsortativity, analogous to Eq.~\eqref{assort}:
\begin{equation}\label{globalr3k}
\bar{r}_{3K}=\sum_{k=2}^\infty p(k)r_{3K}(k)
\end{equation}
% Studying Eq.~\eqref{3kvar} carefully, we see that $\sigma_{3K}^2(k) \ge 0$ implies $1+(k-1)r_{3K}(k)\ge0$, therefore
which, in turn, implies that
\begin{equation}
\bar{r}_{3K} \ge-\sum_{k=2}^\infty p(k) \frac{1}{k-1} \ge -\left\langle \frac{1}{k-1}\right\rangle.
\end{equation}
Negative transsortativity is bounded by the mean of the inverse and is therefore typically small.

\begin{figure}
  \centering
\includegraphics[width=0.95\columnwidth]{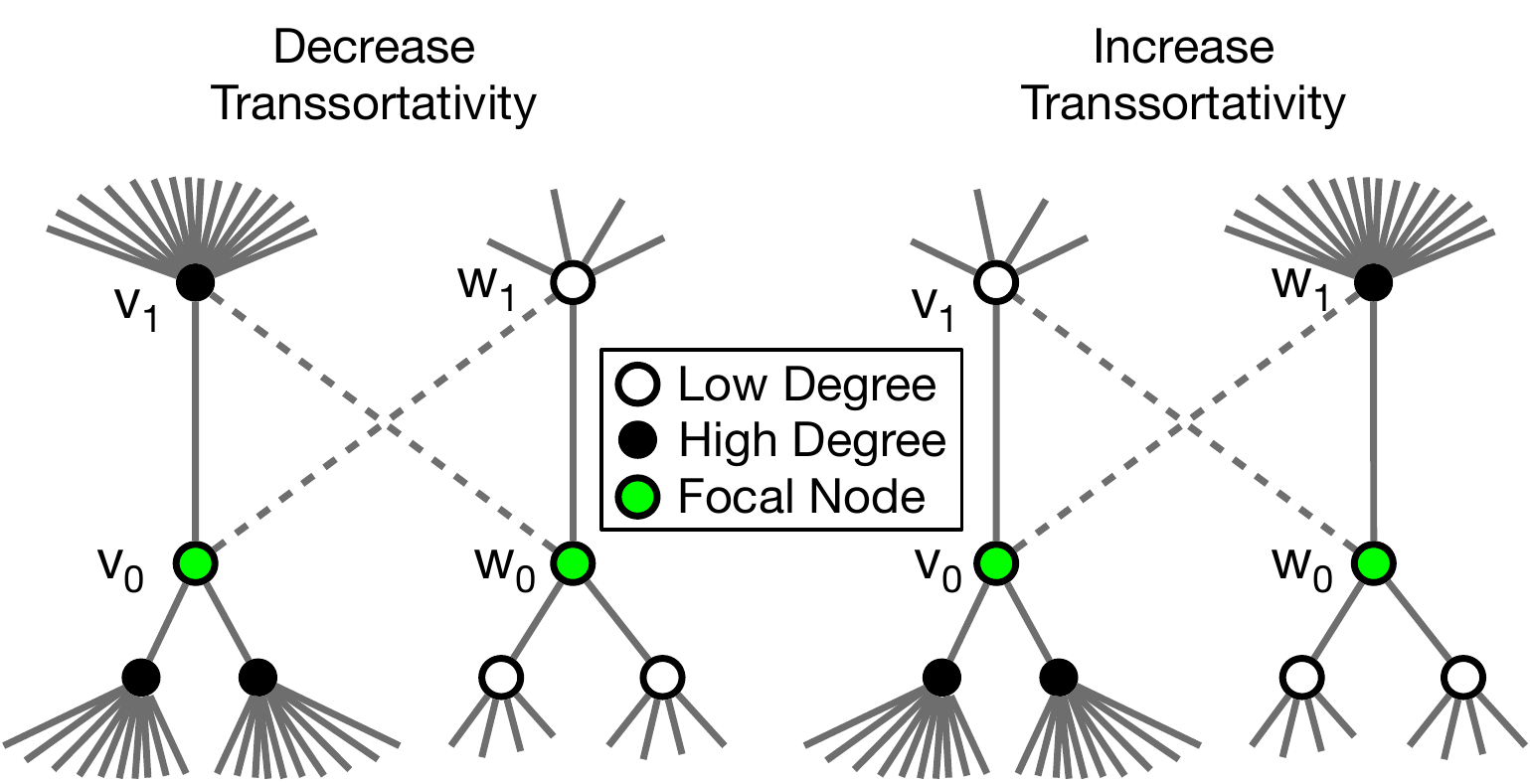}%Karate/CombinedKarateFigs.pdf}
  \caption{
(Color online) Varying mean transsortativity in networks. The algorithm takes two nodes, $v_0$ and $w_0$, that have the same degree, and picks respective neighbors $v_1$ and $w_1$. Left: to reduce transsortativity, $v_1$ and $w_1$ swap edges (\emph{dashed lines}) if this makes neighbor degrees become more diverse. Right: to increase transsortativity, $v_1$ and $w_1$ swap edges if this makes neighbor degrees become more similar. Since $v_0$ and $w_0$ have the same degree, the degree distribution and assortativity remain unchanged.
}\label{fig:algorithm}
\end{figure}

\begin{figure*}
  \centering
\includegraphics[width=0.65\textwidth]{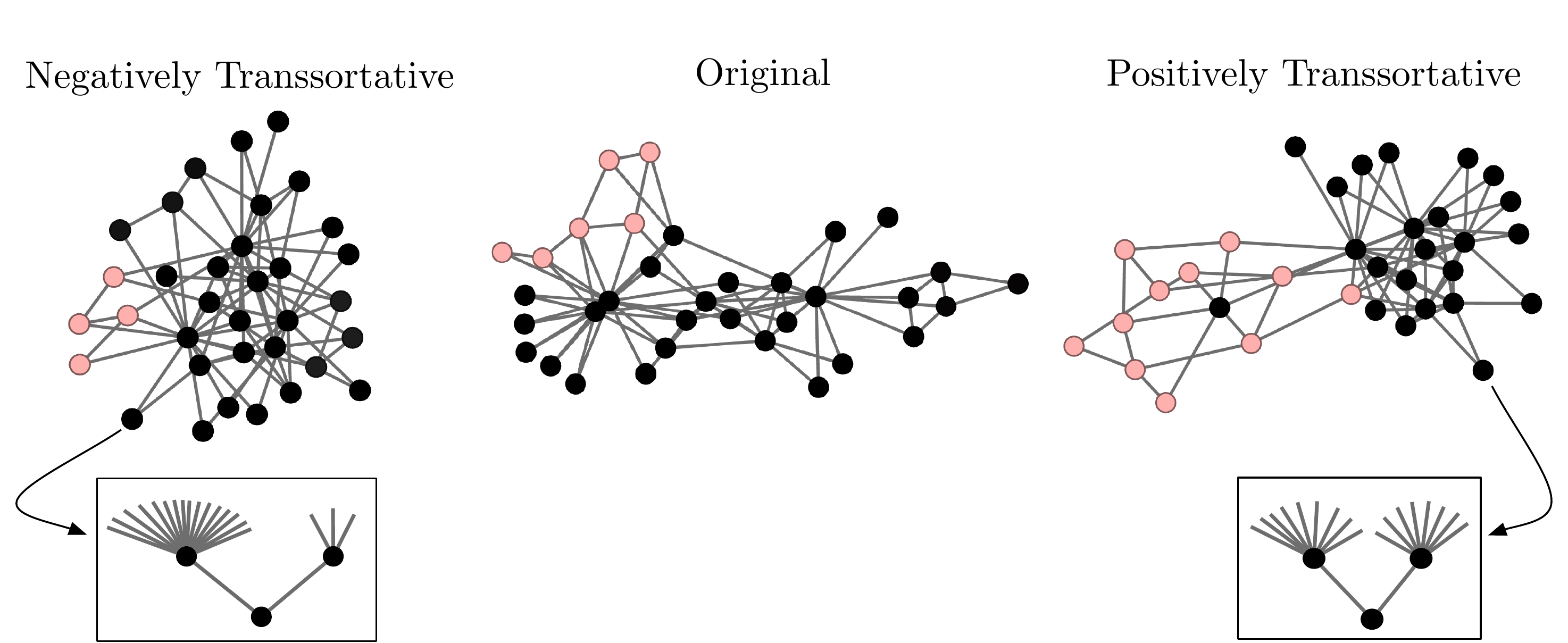}%Karate/CombinedKarateFigs.pdf}
  \caption{Examples of the transsortativity algorithm on Zachary's Karate Club network. Left: negatively transsortative ($\bar{r}_{3K}=-0.40$), center: original ($\bar{r}_{{3K}}=-0.098$), and right: positively transsortative ($\bar{r}_{3K}=0.46$) versions of the network. Lighter dots correspond to the giant vulnerable cluster (GVC) in the Watts model with threshold $\phi = 0.2$ (see cascade discussion in main text). %  Note that the rewiring procedure does not change the degree distribution nor degree assortativity.
   Insets: examples of neighbor degree correlations within the negatively and positively transsortative networks.%(b) Neighbor assortativity versus degree for the rewired and original Karate Club networks.
% from degree specific $r_{\emph{3K}}(k)$ in Zachary's Karate Club network.
}\label{fig:karate viz}
\end{figure*}

\subsection{Transsortativity Rewiring Algorithm}
We use a rewiring algorithm~\cite{Wu2018}  that preserves the degree distribution and degree assortativity %joint degree distributions
(i.e., \emph{1K} and \emph{2K} structure), but can independently vary the transsortativity (\emph{3K} structure), as shown in Fig.~\ref{fig:algorithm}. Two nodes $v_0$ and $w_0$ of equal degree are chosen at random. Let that degree be $k_0$.  One of the $k_0$ neighbors of $v_0$ is chosen at random, denoted $v_1$, and one of the $k_0$ neighbors of $w_0$ is chosen at random, denoted $w_1$. % Two edges, with end nodes $v_0$, $v_1$, $w_0$ and $w_1$ are chosen where one node in each edge has the same degree ($v_0$ and $w_0$ have degree $k_{0}$, shown in Fig.~\ref{fig:algorithm}).
To decrease transsortativity (Fig.~\ref{fig:algorithm} left), edges $\{v_0,v_1\}$ and $\{w_0,w_1\}$ are replaced with edges $\{v_0,w_1\}$ and $\{w_0,v_1\}$ if the edge swap makes $v_0$ and $w_0$ have more diverse neighbor degrees: smaller $r_{3K}(k_{0})$. To increase transsortativity (Fig.~\ref{fig:algorithm} right), the edges are swapped if this makes $v_0$ and $w_0$ have more similar neighbor degrees:~larger $r_{3K}(k_{0})$.

We further illustrate the algorithm on Zachary's Karate club network~\cite{Zachary1977}, shown in Fig.~\ref{fig:karate viz}. This network contains 34 members of a Karate club, with 78 social ties between them. The network is highly disassortative, with $r_{2K}=-0.476$.
Before rewiring, the original Karate club network has neutral transsortativity: $\bar{r}_{3K}=-0.098$ (Fig.~\ref{fig:karate viz} middle). Our rewiring algorithm can create networks with mean transsortativity ranging from % a negative extreme,
$\bar{r}_{3K}=-0.40$ (Fig.~\ref{fig:karate viz} left) to % a positive extreme,
$\bar{r}_{3K}=0.46$ (Fig.~\ref{fig:karate viz} right). While the degree distribution and degree assortativity are identical in all cases, nodes in the negatively transsortative %rewired
network have neighbors with widely varying degree, while nodes in the positively transsortative %rewired
network have neighbors with similar degree (see figure insets), producing very different topologies.

\subsection{Transsortativity and Network Phenomena}

\paragraph{Majority illusion}
We now consider the impact transsortativity has on network phenomena. First, we look at networks where nodes have particular attributes: examples might be gender, political affiliation, or economic status. It has been shown that certain topologies produce a ``majority illusion''~\cite{Lerman2016}, where a significant fraction of nodes observe that a majority of their neighbors have a specific attribute, even when it is relatively uncommon. Transsortativity can amplify the majority illusion. To understand why, consider a hypothetical social network where an individual's popularity correlates with an attribute such as happiness~\cite{Bollen2017}. As a consequence, happier people would be more popular in this network and vice versa. Thus, even if only a small minority of individuals are happy, they would have a tendency to share many neighbors. These neighbors see a large fraction of friends that are happy, and a na\"ive observer would conclude that most of his or her friends are happy.%\note{appear together as neighbors of many individuals people to conclude that most of their friends are happier.}

The following straightforward analysis demonstrates this phenomenon explicitly. Consider a degree-$k$ node with a binary attribute $x\in\{0,1\}$, such as gender or political affiliation, and assume that $x=0$ for a majority of nodes. Let $f(k)$ be the probability that a majority of its $k$ neighbors have attribute value $x'=1$.
The overall probability of majority illusion is
\begin{equation}
P_{>\frac{1}{2}}=\sum_{k=1}^{k_{\max}}p(k)f(k).
\label{eq:mi}
\end{equation}
If the network is locally tree-like, neighbor attributes could simply arise as the outcomes of independent Bernoulli random trials with success probability denoted $\mu_x(k)=P(x'=1|k)$.  Then, since $f(k)$ is the probability of having more than $k/2$ such successes, it could be expressed using a binomial distribution and corresponding Gaussian approximation:
\begin{align}
f(k)&=\sum_{m=\left\lceil\frac{k+1}{2}\right\rceil}^k\binom{k}{m}\mu_x(k)^m[1-\mu_x(k)]^{k-m} \nonumber\\
&\approx 1-\Phi\left[\frac{1-2\mu_x(k)}{2\sigma_x(k)}\right],
% &\approx 1-\Phi\left\{\frac{\frac{1}{k}\left\lceil\frac{k+1}{2}\right\rceil-P(\xpr=1|k)}{\sigma_x(k)}\right\},
\label{fk_gaussian}
\end{align}
where $\Phi$ is the cumulative distribution function of the normal distribution, and $\sigma^2_x(k)=\mu_x(k)[1-\mu_x(k)]/k$ is the variance in the mean neighbor attribute value of a degree-$k$ node.

However, in networks where node attributes are correlated with their degrees, transsortativity leads to correlations between attributes $\xpr_i,\xpr_j$ of pairs of two-hop neighbors. Assuming no higher-order correlations exist, such as among connected subgraphs of four nodes (\emph{4K} structure), it is sufficient to replace the expression for $\sigma^2_x(k)$ by the variance of a \emph{correlated} binomial distribution~\cite{Hisakado2006,Wu2017}. The same calculation as in Eq.\eqref{3kvar} gives
\begin{equation}
\sigma^2_x(k)=\frac{1}{k}\mu_x(k)[1-\mu_x(k)] + \frac{k-1}{k}\,\mathrm{Cov}(\xpr_i,\xpr_j|k).
\label{eq:correlated_binomial}
\end{equation}
Under a simplifying assumption of a bivariate normal distribution for attribute $x$ and degree $k$ (see Supplemental Material),
\begin{equation}
\mathrm{Cov}(\xpr_i,\xpr_j|k)\approx\rho_{kx}^2\frac{\mathrm{Var}(x)}{\mathrm{Var}(k)}\mathrm{Var}(\kpr|k)r_{3K}(k),
\end{equation}
where $\rho_{kx}=\mathrm{Cov}(k,x)/\sqrt{\mathrm{Var}(k)\mathrm{Var}(x)}$ is the degree-attribute correlation.
Then, $\sigma^2_x(k)$ is close to linear in the transsortativity value $r_{3K}(k)$, and it follows from Eq.~\eqref{fk_gaussian} that increasing transsortativity amplifies the majority illusion.  Adopting our earlier analogy, if popular people are happier, a transsortative network structure can create the perception that most people are happier, even when few people are.%In summary, positive transsortativity $r_{3K}(k)$ creates positive neighbor attribute correlation $\rho(\xpr_i,\xpr_j|k)$ (if the network degree distribution is not fat-tailed where Gaussian approximation is applicable), which amplifies the ``majority illusion'' in networks.

\begin{figure}
  \centering
\includegraphics[width=0.95\columnwidth]{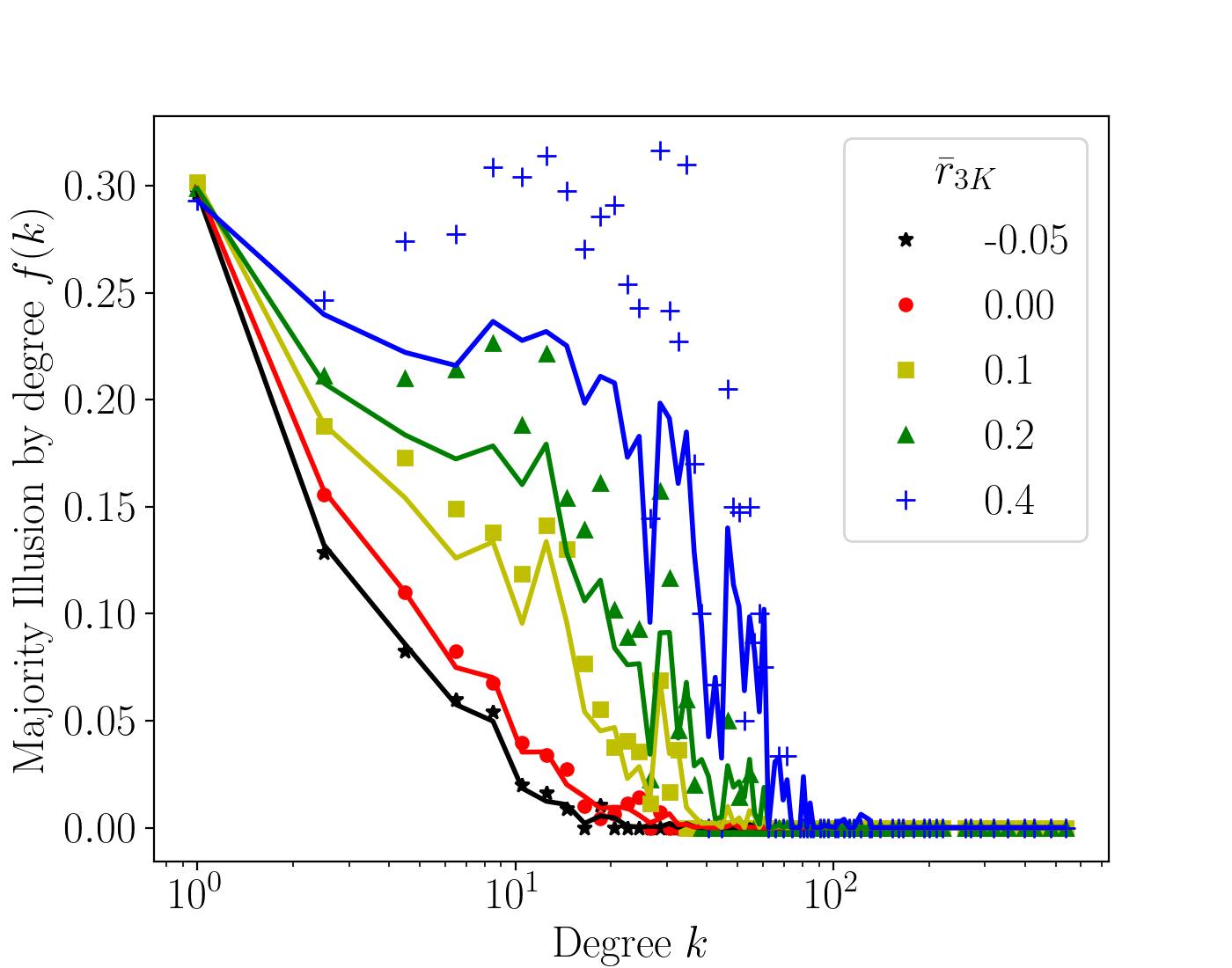}%mi21s_n15_01_transsortative-crop.eps}%{mi21s_n15_01.eps}
  \caption{(Color online) Strength of majority illusion effect on power-law networks from configuration model with 10,000 nodes and exponent $\alpha=2.1$.  Networks are rewired for different mean transsortativity values, both positive and negative.  1\% of nodes have binary attribute value $x=1$, configured to create degree-attribute correlation $\rho_{kx}=0.6$.  Lines show results from binomial model $\eqref{fk_gaussian}$ with measured mean and variance of (correlated) neighbor attribute values.  Symbols show empirical fraction of degree-$k$ nodes for which a majority of neighbors have attribute $x'=1$. %are active (``majority illusion''), and the lines are theoretical predictions by Eq.~\eqref{fk_gaussian} given the \emph{3K} structure, degree-attribute structure, and monophily. %\note{[DONE! In figure, title the legend ``Global Transsortativity'', $\bar{r}_{3K}$'' (split title into 2 lines) and remove ``$r_{3K}=$''.]}
}
\label{fig:mi}
\end{figure}

We demonstrate this effect in Fig.~\ref{fig:mi}, on power-law networks (exponent $\alpha=2.1$, degree assortativity $r_{2K}=-0.15$), with degree-attribute correlation $\rho_{kx} = 0.6$, rewired to vary mean transsortativity $\bar{r}_{3K}$ (from $-0.05$ to $0.4$).  Only 1\% of the nodes have attribute value $x=1$, while the rest have attribute $x=0$. We show the results for $f(k)$ from an exact calculation for $k=1,2$ and the normal approximation in Eq.~\eqref{fk_gaussian} for $k\ge 3$, based on the measured values of $\mu_x(k)$ and $\sigma_x(k)$.  (See Supplemental Materials for details and for results on differently generated networks.)  We also plot the empirically measured fraction of degree-$k$ nodes that experience the majority illusion.  In both cases, the majority illusion effect grows significantly with increasing transsortativity $\bar{r}_{3K}$: for moderate degree $k$, the fraction of nodes that see the 1\% minority as being a majority in their neighborhoods can be an order of magnitude larger at $\bar{r}_{3K}=0.4$ than at $\bar{r}_{3K}=0$.  Furthermore, the model results are qualitatively consistent with the empirical results, suggesting that the tree-like approximation is justified and that degree correlations beyond transsortativity do not play an important role.

\paragraph{Overdispersion}
While other mechanisms have been proposed for introducing correlations between neighbor attributes, their consequences are more limited.  In the field of social networks, the phenomenon of \emph{overdispersion} refers to cases where the attribute variance $\sigma_x^2(k)$ is larger than a simple binomial model would predict.  This is associated with a segregation effect where nodes are unexpectedly likely or unlikely to have neighbors possessing the attribute.  Empirical studies have suggested that overdispersion can occur when the neighbor attribute probability $\mu_x(k)$ itself varies from one node to another~\cite{Zheng2006}, and moreover that this can induce pair correlations between neighbors~\cite{Altenburger2018}.  Indeed, from the law of total covariance, one may show (see Supplemental Material) that $\mathrm{Cov}(\xpr_i,\xpr_j|k)$ is simply equal to the variance of the quantity $\mu_x(k)$.  However, such an approach only accounts for positive neighbor correlations and a resulting increase in $\sigma_x^2(k)$ (see Eq.~\eqref{eq:correlated_binomial}).  By instead understanding these effects as a consequence of transsortativity, we arrive at an explanation that simultaneously includes positive and negative attribute correlations, overdispersion as well as underdispersion, and segregation of neighbor attributes.

% \subsection{Threshold Cascades}
\paragraph{Global cascades}
Finally, we demonstrate how a network's transsortative structure can significantly alter dynamics of phenomena unfolding on it. We consider the popular Watts threshold model describing cascade dynamics~\cite{Watts2002}, where nodes can be either ``active'' or ``inactive.''  Starting from a single active seed, nodes in the network become activated whenever more than a given %(greater than some threshold)
fraction $\phi$ of their neighbors are active. This model has been used to describe contagion processes as well as the spread of ideas and opinions spread in social networks~\cite{Centola2010,Centola07b}.

\begin{figure}
  \centering
\includegraphics[width=0.9\columnwidth]{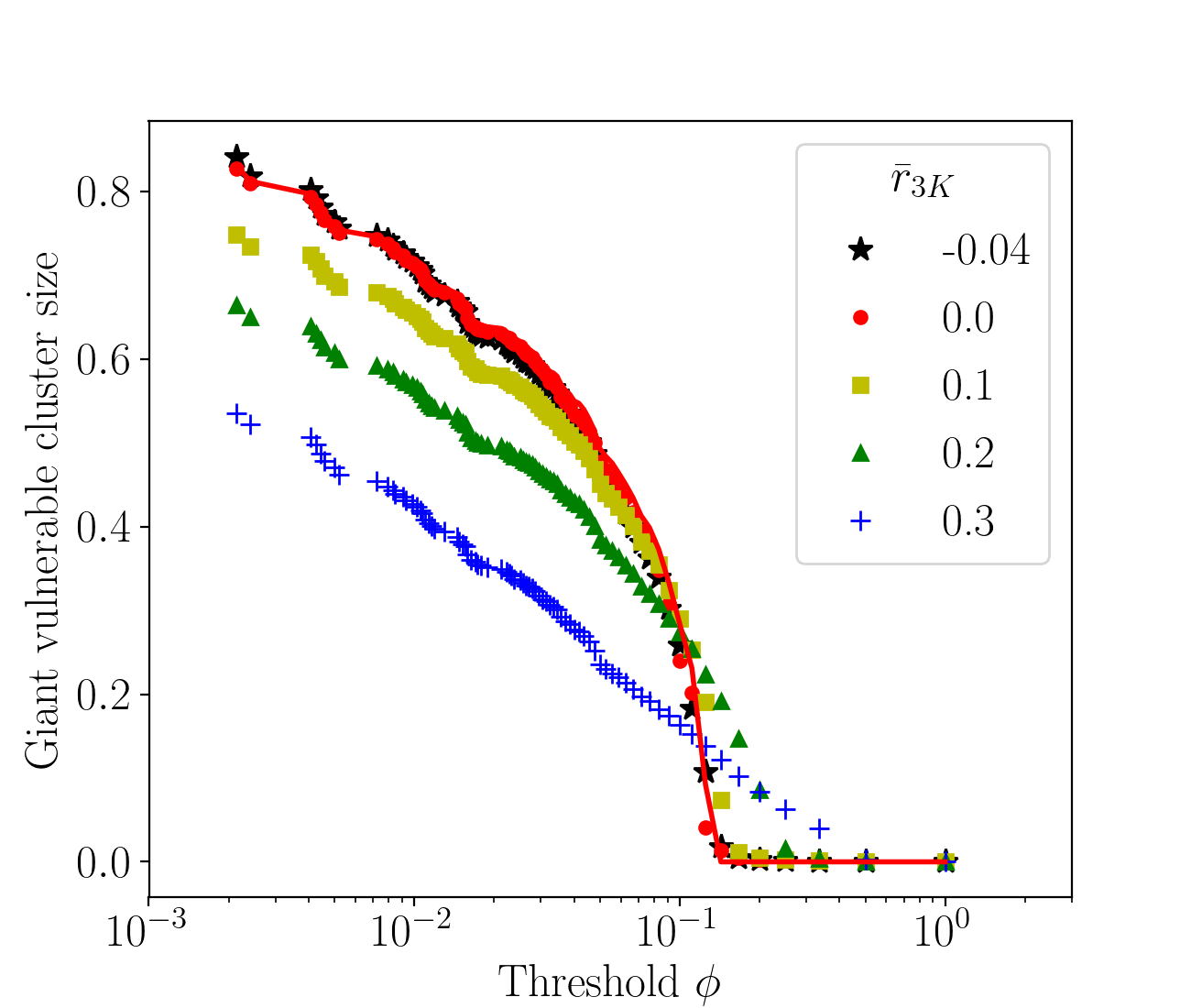}%power24_nbassort_transsortative-crop.eps}%{power24_nbassort.eps}
  \caption{(Color online) Cascades triggered by a single active node, on power-law networks from configuration model with 10,000 nodes, exponent $\alpha=2.4$, and degree assortativity value $r_{2K}=-0.07$.  Solid line shows % predicted results from the generating function formulation in~\cite{Wu2018} for baseline
theoretical results for baseline case $\bar{r}_{3K}=0$: below a critical threshold value of $\phi^*=1/7$, a finite fraction of nodes belongs to the GVC.  Symbols show simulated results on networks rewired for different mean transsortativity values.
}\label{fig:power24_nbassort}
\end{figure}

In the Watts model, global cascades occur when the required threshold $\phi$ is below a critical value $\phi^*$: the largest cascade, known as the giant vulnerable cluster (GVC), then extends to a finite fraction of the network~\cite{Dodds2009,Payne2009,Wu2018}.  Figure~\ref{fig:power24_nbassort} illustrates how $\phi^*$ varies when networks are rewired for different transsortativity values.  Increasing transsortativity tends to increase the critical threshold for the GVC, from the value $\phi^*=1/7$ predicted by the generating function formulation in~\cite{Wu2018} for $\bar{r}_{3K}=0$,
% theoretical baseline value of $\phi^*=1/7$ for $\bar{r}_{3K}=0$
to $\phi^*=1/2$ for $\bar{r}_{3K}=0.3$.  Just as transsortativity amplifies the majority illusion effect in low-to-moderate degree nodes, it can cause nodes to perceive a small fraction of active nodes as a large fraction of their neighbors, and  become activated themselves.  Thus, even moderate transsortativity can have a significant impact on the formation of global cascades.

For the special case of $\phi=1/2$, active nodes are in fact precisely those experiencing the majority illusion, so transsortativity has the direct effect of amplifying cascade size. %, thus confirming our intuition for that particular parameter.
% Because the majority illusion is amplified by transsortativity, so are cascade sizes.
Figure~\ref{fig:power24_nbassort} demonstrates that this effect occurs generally for large enough values of $\phi$, since in that regime the cascade is sparse and spreads as a branching process; the (correlated) binomial model for the majority illusion applies here.
A further example is seen in Fig.~\ref{fig:karate viz}, where the GVC (at $\phi=0.2$) nearly doubles in size from the original network to the positively transsortative network.  However, for smaller values of $\phi$, cascades spread far more densely~\cite{Dodds2009}.  The locally tree-like approximation of the network no longer provides a valid description of neighbor activity, and Fig.~\ref{fig:power24_nbassort} shows that increasing transsortativity suppresses rather than amplifies the GVC there.  It remains an open question whether this could in part be due to a coarsening effect, where attribute segregation results in the formation of domains in the network that impede the growth of the GVC. An analogous effect has been noted in~\cite{Newman2002Assort,Payne2009} under increasing degree assortativity.

\section{Discussion}
We have defined transsortativity in a network as the (two-hop) degree correlation between a pair of neighbors of a node, by analogy to degree assortativity, which represents the (one-hop) degree correlation between a node and its neighbor~\cite{Newman2002Assort}. Transsortative structure has a significant impact on perceptions and phenomena in the network. It can significantly amplify the majority illusion effect, % by as much as 37.6\%,
and increase the critical threshold for global cascades in the Watts threshold model by more than three-fold.  Moreover, transsortativity explains overdispersion in network neighborhoods, partitioning the network into domains where unexpectedly high or low concentrations of an attribute are observed~\cite{Zheng2006,Altenburger2018}.  In real networks, both positive and negative transsortativity occur; in synthetic networks, we show how to increase or decrease transsortativity while preserving lower-order network statistics such as degree distribution and assortativity.  Our work explains how well-established simulation methods, such as configuration models~\cite{Molloy1995,Hofstad2016} and degree-preserving rewiring algorithms~\cite{Noldus2015} do not fully capture how real-world networks affect network phenomena.%, may need to incorporate transsortative structure in order to account for realistic network structure.% allow accurate predictions of attribute dynamics on networks.

This paper raises a number of questions to be addressed by future work. Finite size effects are known to constrain maximum degree and assortativity in scale-free networks~\cite{cutoff}, but the impact of any structural cutoff on transsortativity remains to be studied. Another interesting question is how transsortativity affects evolution of networks. It is conceivable, for example, that transsoratativity and triadic closure jointly increase assortativity in growing networks. Finally, our approach could be generalized to still higher-order structures, for example, connected subgraphs of four nodes (i.e., $4K$ structure), in cases where such expanded statistical models of networks are required. %Greater realistic assumptions may help us better understand why the majority illusion differs significantly from theory for high $\bar{r}_{3K}$.

%\bibliography{references}

\section{Acknowledgements}
{\bf Funding:} This work was supported in part by the Army Research Office (Grant \#W911NF-16-1-0306), whose support is gratefully acknowledged. {\bf Author contributions:} X.-Z.W., A.P., K.B., and K.L. conceived of the research. X.-Z.W. created theoretical and data analysis. X.-Z.W., A.P., K.B., and K.L. discussed results and wrote the
manuscript. {\bf Competing interests:} The authors declare that they have no competing interests.
{\bf Data and materials availability:} Code can be accessed in the following Git repository http://github.io/KeithBurghardt/transsortativity\_code.

\end{document}